\documentclass[smallcondensed,final]{svjour3} 
\usepackage{amssymb,amsmath} 
\usepackage{mathptmx} 
\usepackage{graphicx} 
\usepackage{acronym} 
\usepackage[utf8]{inputenc} 
\usepackage{color}

\begin{document}

\title{Do solids flow?}

\author{F. Sausset \and G. Biroli \and J. Kurchan}

\institute{ F. Sausset \and G. Biroli \at Institut de Physique Th\'eorique, CEA, IPhT, F-91191 Gif-sur-Yvette, France CNRS, URA 2306. \\\email{sausset@lptmc.jussieu.fr} \\\email{giulio.biroli@cea.fr} \and J. Kurchan \at Physique et Mécanique des Milieux Hétérogènes, PMMH, ESPCI, 10 rue Vauquelin, Paris, France 75005, CNRS, UMR 7636 \\\email{jorge.kurchan@espci.fr}} 

\date{\today}

\maketitle
\begin{abstract}
	Are solids intrinsically different from liquids? Must a finite stress be applied in order to induce flow? Or, instead, do all solids only look rigid on some finite timescales and eventually flow if an infinitesimal shear stress is applied? Surprisingly, these simple questions are a matter of debate and definite answers are still lacking. Here we show that solidity is only a time-scale dependent notion: equilibrium states of matter that break spontaneously translation invariance, e.g. crystals, flow if even an infinitesimal stress is applied. However, they do so in a way inherently different from ordinary liquids since their viscosity diverges for vanishing shear stress with an essential singularity. We find an ultra-slow decrease of the shear stress as a function of the shear rate, which explains the apparent yield stress identified in rheological flow curves. Furthermore, we suggest that an alternating shear of frequency $\omega$ and amplitude $\gamma$ should lead to a dynamic phase transition line in the ($\omega$,$\gamma$) plane, from a 'flowing' to a 'non-flowing' phase. Finally, we apply our results to crystals, show the corresponding microscopic process leading to flow and discuss possible experimental investigations. 
\end{abstract}

\section*{ } 

\label{sec:_}

We clearly know how to make the difference between a solid and a liquid in everyday life. However, the question of whether solid and liquid are distinct states of matter is far from being conclusively answered in the literature. A difficulty in addressing this issue is that even ordinary liquids may look solid, as one learns when diving into the water from an height of, say, 10 meters without being a pro. Actually, supercooled liquids and soft matter, such as emulsions, foams, and pastes behave like bona-fide solids, if probed on timescales that range from microseconds to hours. This phenomenon, called visco-elasticity, is a property of most liquids: they respond as elastic solids on short timescales and as a real liquid on long timescales (for an amusing illustration see the pitch experiment \cite{pitchexperiment}). Of course the meaning of long and short depends on the system at hand: the viscous-elastic crossover takes place on timescales of the order of $10^{-12}$ s for water \footnote{Therefore, in the diving example water 'looks' solid but not because of visco-elasticity. See \cite{splash} for a detailed explanation of the physics of splash.} and on the order of hours for supercooled glycerol close to $190K$. The question we address in this work is whether {\it all known solids} are actually like that, i.e. liquid-like on large enough timescales. This is the conjecture of some rheologists, as nicely argued in \cite{pantarei}. On the other hand, general theoretical results, see e.g. \cite{chaikinlubensky}, argue that rigidity is a consequence of breaking translation invariance (in all directions): states of matter characterized by this spontaneous symmetry breaking, like crystals, must be rigid and have a finite elastic shear modulus. 

How can we reconcile these two apparently antitethic claims? First, it is important to remark that solids in nature contain often a large quantity of non-equilibrium defects that allow them to flow rather easily. The main example is provided by dislocations in crystals. They cost infinite energy \footnote{The energy of a single dislocation diverges so strongly in the thermodynamic limit for dimension higher than 2 that the equilibrium dislocation density is zero \cite{chaikinlubensky}.} and, hence, they are absent in three-dimensional crystals at thermal equilibrium. Although equilibrium crystals can be prepared \cite{balibar,silicon}, typical growth procedures lead to imperfect crystals characterized by a dislocation density that ranges from $10^6-10^{10} cm^{-2}$. When a shear stress $\sigma$ is applied, dislocations move with a velocity proportional to the stress and lead to what is called plastic flow \cite{hirth}. The experimental examples studied in \cite{pantarei} are probably all characterized by some kind of non-equilibrium defects. On the other hand, the theoretical results cited in \cite{chaikinlubensky} refer to {\it equilibrium} states of matter that break completely translation invariance, which we will call in the following "perfect solids". One possibility would be therefore that perfect solids are rigid but that non-equilibrium ones flow. Instead, here we show that even perfect solids do flow on long enough timescales if an infinitesimal macroscopic shear stress is applied. Despite this property, we also show that perfect solids are different from ordinary liquids because their viscosity diverges for vanishing shear stress with an essential singularity---contrary to the finite viscosity characteristic of ordinary liquids. The role of the thermodynamic limit is a subtle and important issue. In fact, it is clear that any {\it finite} system cannot be rigid and will flow if one waits a long enough time, see e.g. \cite{anderson}. The crucial question is whether or not this time scales with the system size. In the former case real solids would flow but this would happen on timescales too large to be relevant for any practical purposes, maybe even larger than the age of the universe. We shall show that instead the timescales for flow are long but finite in the thermodynamic limit and attainable in experiments. 
\begin{figure}
	[htbp] \centering 
	\includegraphics[width=6cm]{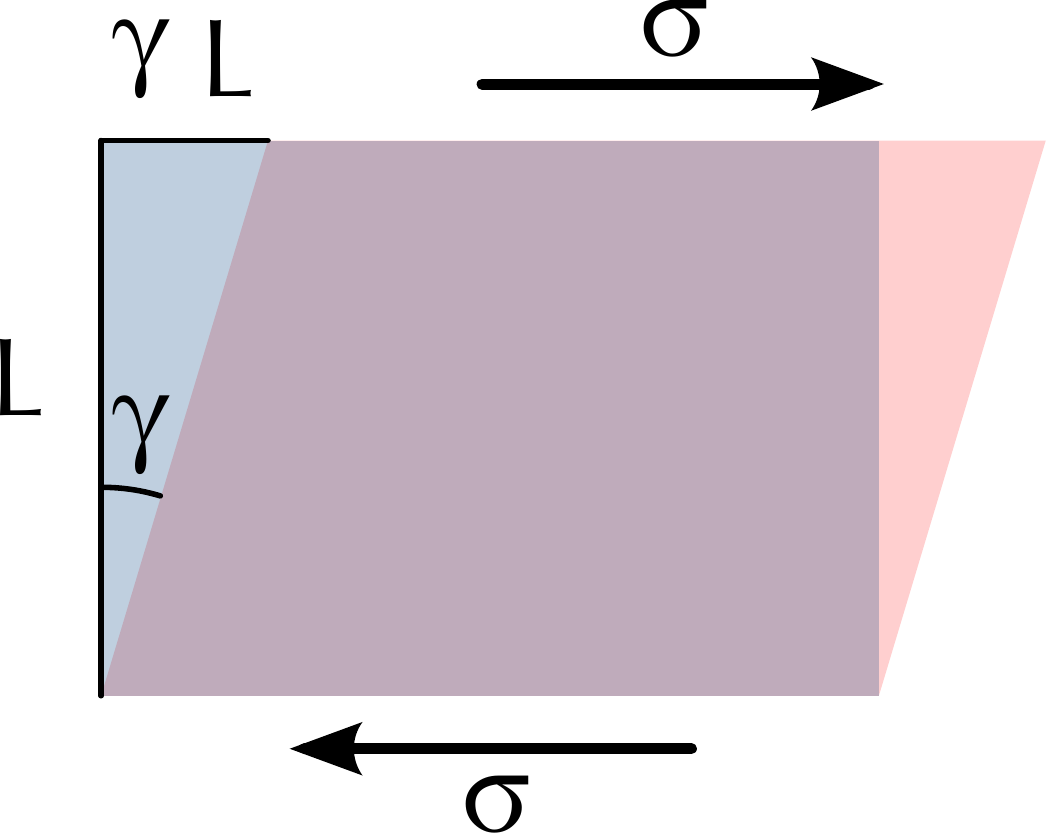} \caption{(Color online) Schematic representation of the deformation due to a shear stress. The amount of deformation is measured by the angle $\gamma$ which is proportional to the shear stress $\sigma$. The proportionality constant is the inverse of the elastic shear modulus $G$.} \label{fig:sketch} 
\end{figure}

Our starting point is similar to the usual nucleation arguments for metastable phases. Consider a solid of size $L^d$ (where $d$ is the spatial dimension), --- it could be 
an equilibrium phase such as a crystal, a quasi-crystal or an ideal glass, if such a thing exists; or even a metastable state such as a crystal with many entangled dislocations. We  apply a very small but finite shear stress $\sigma$, see Fig. \ref{fig:sketch}. After a very rapid equilibration time the system will deform by an amount $\gamma=\sigma/G_P$ where $G_P$ is what is called in practical applications the elastic shear modulus.
This deformation leads to an increase in the free energy of the system that according to elasticity theory is given by: $L^d \frac{\sigma^2}{2G_P}$. This implies that the deformed solid has a free energy {\it extensively} higher than the one of the equilibrium undeformed state and, as a result, that the deformed solid is metastable and has a finite life-time, except if the free energy barrier to escape from this metastable equilibrium is infinite -- which is not the case, as we shall show in the following.
\begin{figure}
	[htbp] \centering 
	\includegraphics[width=9cm]{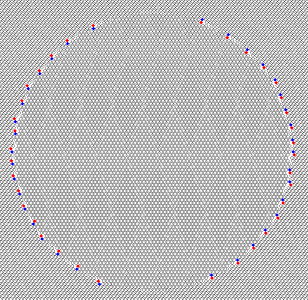} 
\caption{Defects structure around a relaxed cavity embedded in a sheared system. The gedanken experiment discussed in the main text consists in reshuffling all particles inside the cavity from the deformed to the undeformed configuration. Here we show the case of a two-dimensionnal hexagonal crystal for a strain $\gamma = 0.4$. The interaction between particles is of hard core type with a radius $0.35a$, where $a$ is the lattice spacing. Particles are plotted as disks with radius $a/2$. White (empty) particles have six neighbours and thus satisfy the hexagonal order, whereas red and blue ones correspond respectively to particles having seven and five neighbours, which are elementary topological defects called disclinations. Note that a 5-7 dipole of disclinations corresponds to a dislocation, the Burgers vector of which is perpendicular to the dipole axis. Here, the cavity has a radius bigger than $R^*$, the critical radius of a relaxed nucleus in a such system, see text.} \label{fig:cavity} 
\end{figure}

The introduction of a `static' modulus begs an explanation, since we shall argue in what follows that even perfect  solids 
flow under finite stress. We shall show that under the effect of a stress, the system responds in two time-steps: a rapid one corresponding to the elastic response inside the metastable deformed state,
followed by a slow flow  on  large timescales. The value $G_P(t)$ we use here is defined
 as the elastic shear modulus obtained at times long compared to the elastic deformation, but short compared with the stress-dependent nucleation time~\footnote{This is in strict analogy with the susceptibility in a ferromagnet: a small field applied against the magnetization will induce a linear response at intermediate times, but will eventually
 destabilize the phase.}. The two-timescale response of course applies (by definition) to viscoelastic systems, but we shall argue here that it holds also  for true solids 
 at long enough timescales: the difference with viscoelasticity being that for a solid the long-time flow rate vanishes faster than $\sigma$ at  small stresses, thus implying an infinite viscosity in the limit of zero stress. 

Let us now introduce our main argument by focusing on a cavity of radius $R$ inside the deformed solid. Imagine reshuffling all particles inside the cavity from the deformed to the undeformed configuration. For example, in the crystal case this is shown in Fig.~\ref{fig:cavity}. Then, while still maintaining the applied shear stress $\sigma$, we let the solid evolve dynamically from this configuration. After a short equilibration time the system reaches a metastable equilibrium. The free energy difference between this state and that without cavity reads: 
\begin{equation}
	\label{eq:nucl} \Delta F(R)=R^{d-1}\Omega(\sigma)-R^d\frac{\sigma^2}{2G_P} 
\end{equation}
where the first term corresponds to the free energy cost one has to pay due to the mismatch at the surface of the cavity between deformed and undeformed state. $\Omega(\sigma)$ is like a surface tension and can in principle depend on $\sigma$. For small $\sigma$, the limit we are interested in, the dependence on $\sigma$ matters only if $\Omega(\sigma)$ does not go to a constant and vanishes \footnote{We consider particles interacting with a short range non singular potential, therefore it is sure that $\Omega(\sigma)$ has to be bound by a finite positive constant for $\sigma \rightarrow 0$.}. We will see that this is indeed what happens for crystals. 

$\Delta F(R)$ first increases as a function of R, has a maximum at $R^*=2\frac{d-1}{d}\frac{G_P\Omega(\sigma)}{\sigma^2}$ and then decreases for $R>R^*$. This implies that the nucleation of a droplet of the undeformed solid of size $R$ is unfavorable for radii less than the critical value $R^*$. In contrast nuclei larger than the critical one of size $R^*$ will grow since they are free-energetically favored \cite{Nucleationbookorreview}. Thus, the free-energy barrier for nucleation via a spherical droplet is given by $\Delta F(R^*)=\frac{\sigma^2}{2G_P(d-1)}(R^*)^d$, and this is clearly an upper bound. It is necessarily finite for any finite $\sigma$, although it diverges for $\sigma \rightarrow 0$, as is usual in metastability problems (here $\sigma$ plays the role of the external field inducing metastability). Depending on the value of $\sigma$ and system parameters, either different nuclei are already present in the system with a density $\rho_n\simeq 1/(R^*)^d \exp(-\Delta F(R^*)/k_BT)$ and grow with a finite speed or there are none and one has to wait a time of the order $\tau_0 \exp(\Delta F(R^*)/k_BT)$ to nucleate one that then will grow and invade the system. In both cases, for very small $\sigma$, the characteristic timescale for destabilizing the metastable deformed solid is given by $\tau^*=\tau_0 \exp(\Delta F(R^*)/k_BT)$ where $\tau_0$ is a microscopic timescale that will be discussed later.

As a consequence, after a time of the order $\tau^*$, the deformed solid goes back to the undeformed state. If, however, the external shear stress $\sigma$ is continuously applied, the solid deforms again and the entire process repeats. Although a more refined analysis of the flow process is needed (see below for a first attempt), these simple arguments already suggest that the solid deforms with a shear rate 
\begin{equation}
	\label{flow1} \dot{\gamma}=\frac{\gamma}{\tau^*}=\frac{\sigma}{G_P\tau^*}=\frac{\sigma}{G_P\tau_0}\exp\left( -c\left(\frac{\sigma_y}{\sigma}\right)^{2d-2}\right) 
\end{equation}
where $\sigma_y(\sigma)=G_P\left(\left(\frac{\Omega(\sigma)}{G_P} \right)^d \frac{G_P}{k_B T}\right)^{\frac{1}{2d-2}}$ is an effective scale for the stress and $c$ is a constant, which equals $\pi/3$ in three dimensions. If $\Omega(\sigma)$ does not vanish for $\sigma \rightarrow 0$ neither does $\sigma_y$; if instead $\Omega(\sigma)$ vanishes, then the previous law connecting the shear rate $\dot{\gamma}$ to the shear stress $\sigma$ changes and the exponent $2d-2$ is renormalized. In any case, as long as $\sigma_y$ vanishes less rapidly than linearly in the stress (for crystals we will show that $\sigma_y\propto\sigma^{3/4}(-\log (\sigma/G_P))^{3/4}$) the previous equation leads to an essential singularity for the viscosity of the system: 
\begin{equation}
	\label{flow2} \eta=\frac{\sigma}{\dot{\gamma}}=G_P\tau_0\exp\left(c\left( \frac{\sigma_y}{\sigma}\right)^{2d-2}\right) 
\end{equation}
This is the main result of this work: a perfect solid does flow but with a viscosity that diverges extremely fast when $\sigma \rightarrow 0$. The conclusion of the previous arguments is that what we call in practical applications ``elastic shear modulus", i.e. $G_P$, is, as described above,  only an effective timescale-dependent quantity corresponding to the value of the very long (but not extending until zero frequency) plateau displayed by $G(\omega)$~\cite{rheobook}. Similarly to visco-elastic liquids, perfect ``solids" behave elastically on timescales much less than $\tau^*$ and eventually flow on timescales larger than $\tau^*$. However, the flow properties for small shear stress are very different: instead of the singular relation (\ref{flow1}), ordinary liquids are characterized by a linear Newtonian relation leading to a finite viscosity at $\sigma=0$.

Let us  note that the metastability of the stressed state is closely analogous to the metastability of current-carrying states in superconductors and superfluids \cite{Langer}, and of loaded states in solids \cite{Sethna}. Indeed, the nucleation argument applies to all situations where a system breaks  a continuous symmetry, and acquires a rigidity with respect to long-wavelength deformations along this symmetry
 \cite{chaikinlubensky}: whenever the physical situation is such that   the ``twist" along the zero-modes per unit  length is finite
 in the broken symmetry state, the phase becomes metastable. Such is the case for superfluids, superconductors, solids and smectics. In the context of two-dimensional solids our argument is very similar to the one already developed in \cite{Hal1,oscill} in order to obtain estimates of the viscosity in the strong strain regime. Our purpose is, however, quite different: we give a weaker argument that just gives us an upper (finite) bound for the viscosity, but such that it may be applied {\em in any dimension}, and without invoking the nature of the order responsible for solidity. This allows us to include (putative) ideal glasses whose spatial order is poorly understood at present. It is reassuring that in the case of two dimensional solids, where a detailed analysis can be performed~\cite{Hal1}, our upper bound is just slightly sub-optimal \footnote{Our argument gives $\log \tau \propto -\log \dot \gamma \propto \frac{[\log (\sigma/G_P)]^2}{k_BT}$ in the limit of small shear stress, whereas the true behavior is $\log \dot \gamma \propto [\log (\sigma/G_P)]$ and corresponds to an unbinding mechanism of the dipoles of dislocations that are present in the system in the low temperature crystalline phase.}.

At this stage, it is important to address an apparent paradox: there are rigorous expressions relating the elastic shear modulus to correlation functions evaluated in the equilibrium (unstressed) solid \cite{JSTAT,Evans}. These are used in simulations to obtain quantitative values of the elastic moduli and were shown to be different from zero. There is actually no contradiction with our results, but the reason is subtle and worth discussing in detail. These equilibrium expressions of the shear modulus are obtained via linear response and a fluctuation dissipation relation expressing the response of a crystal to a shear deformation in terms of the correlation functions of the undeformed crystal. This linear response is justified only to the extent that the stress does not induce a macroscopic change, and this will happen either if strain is microscopic $(\gamma$ not of order one), or if it acts for a time $\tau \lesssim \tau^*$ only. Either cases yield a non-zero modulus $G_P$ within the deformed metastable state. 
As for the general theoretical results, cited in the introduction \cite{chaikinlubensky}, stating that the existence of a shear modulus is a consequence of breaking translation invariance in all directions, a careful analysis shows that the breaking of translational invariance implies the finiteness of  
 $G_P$ (defined for times such that the system stays within the metastable non-flowing state),   and not of the true   shear modulus associated with arbitrarily long  times and finite stress. Spontaneous thermal fluctuations
 will not make a true solid flow -- hence the broken translational invariance, but  an external strain of the order of the system size will. Again, this is in strict analogy with a ferromagnet, where spontaneous magnetization fluctuations will not destabilize a phase, but a finite external magnetic field will.

The argument above not only implies that nucleation will make an equilibrium solid under finite stress start flowing, 
but also that it will preclude it from stopping.
In the case of a crystal, as we shall discuss below, the flow is facilitated  initially by dislocations that are nucleated. As is well known,
the dislocations evolve and may entangle, and their motion become restrained.  However, the out of equilibrium phase of interacting 
dislocations cannot sustain a finite stress forever, since the nucleation of an unstressed phase is also possible  in that case.  
What one may observe in such a situation is a first regime in which the defects proliferate and entangle, and a second regime, with much
slower flow, in which the system responds in a manner more similar to that of an amorphous solid.
Needless to say, this also applies to a situation in which there where some nonequilibrium defects present in the initial configuration.

We now apply the previous general arguments to the initial flow of perfect crystals. In order to find the $\sigma$ dependence of $\Omega(\sigma)$ we consider a cylindrical cavity of height and radius equal to $l=\frac a \gamma=\frac{a G_P}{\sigma}$ and whose symmetry axis is perpendicular to the shear direction ($a$ is the lattice spacing). In this case, our construction boils down to creating a single dislocation loop with radius $a/\gamma$ and Burgers vector of length $a$. Both the loop and the Burgers vector are parallel to the basis of the cylinder. Creating the dislocation loop has an energy cost $\frac{2\pi l G_Pa^2}{4\pi(1-\nu)}[\log(l/a)-A]$ where $A$ is a constant of the order of one and $\nu$ is the Poisson ratio \cite{hirth}. However, stress is also released, which leads to an energy gain proportional to $\frac{a \sigma l^2}{2}=\frac{\sigma^2}{2G_P} l^3$ \cite{hirth}. Thus, on the scale $l=\frac a \gamma$, we have found a microscopic expression for the energy gain and energy loss terms introduced previously. By equating the generic expression $\Omega(\sigma) l^2$ to its value for $l=\frac a \gamma$ we find $\Omega({\sigma})=\frac{ca\sigma}{4\pi(1-\nu)}[-\log (\sigma/G_P)-A]$ where $c$ is a constant, whose value depends on the geometry of the cavity and which we will take equal to one in the following. Assuming $\nu\simeq1/3$, which is a typical value for solids, one obtains the following laws for crystals in the limit of very small $\sigma$: 
\begin{equation}
	\label{crystal-tau} \tau^*=\tau_0 e^{k\frac{\sigma_y^c\left(-\frac 1 8 \log \frac{\sigma}{\sigma_y^c} \right)^3}{\sigma}}, \quad \eta=G_P\tau_0 e^{k\frac{\sigma_y^c\left(-\frac 1 8 \log \frac{\sigma}{\sigma_y^c} \right)^3}{\sigma}}, \qquad with \quad \sigma_y^c=\frac{G_P^2 a^3}{k_BT}, 
\end{equation}
where $k$ is a constant whose value is close to one and is system-dependent; $\sigma_y^c$ is an effective stress scale. Note that since generically $G_P\simeq k_BT/a^3$ \cite{chaikinlubensky}, the typical value of $\sigma_y^c$ is of the order of $G_P$ (or slightly larger). The value of $R^*$ for crystals turns out to be of the order of $-a\log(\sigma/G_P)/(\sigma/G_P)$. Thus, roughly, the critical nucleus is formed by a few dislocation loops, whose nucleation has already been discussed in the literature, see e.g. \cite{hirth,cohen}. This also allows one to estimate the microscopic timescale. Whether the critical nuclei are already present in the system or one has to be nucleated, $\tau_0$ should be roughly given by the time to nucleate a kink-antikink pair in a dislocation line (in 3D), which is of the order of the Debye time multiplied by $\exp(2W/k_BT)$ where $W$ is the kink energy. This is in fact the typical time to nucleate a microscopic dislocation loop and also the time on which a dislocation moves of one lattice spacing.

In the previous arguments, the non-equilibrium steady flowing state was described in a very phenomenological way. In the following we investigate in more detail the flow process and endeavor to obtain some predictions for the flow curve of a "solid". For not too small shear stresses, there is a finite density of critical nuclei \footnote{From a purely statistical mechanics point of view, if one considers an infinite system then there is always a finite density of critical nuclei.} and one has to take into account their growth and their continuous deformation due to a constant shear rate. Forgetting the continuous deformation for the time being, one finds that the growth is a two-stage process: critical nuclei first expand driven by a free-energy gain, until they start to overlap. At this point the growth becomes a coarsening-like process \cite{Brayreview}. Since this is expected to be much slower, we will neglect the latter in the following. We model the system as divided in mesoscopic blocks characterized by a certain value of the stress $\sigma$. The block sizes are of the order of the distance between critical nuclei. We assume that the stress inside the block increases with time due to the constant shear rate. A critical nucleus can appear inside the block with a rate $1/\tau^*(\sigma)$ (see eq. (\ref{flow1})). This leads to a subsequent decrease in the stress that we model with a transition $\sigma \rightarrow 0$ \footnote{Clearly this is not an instantaneous process and a more refined model should account for the finite time it takes to the nucleus to growth and invade the block.}. The equation for the evolution of the distribution of block stresses, $P(\sigma,t)$, is therefore: 
\begin{equation}
	\label{eq:stressEvolution} 
	\partial_t P(\sigma,t) = -G_P \dot{\gamma} \, 
	\partial_\sigma P(\sigma,t) - \frac{1}{\tau^*(\sigma)} \, P(\sigma,t)+\delta(\sigma)\int_0^{+\infty}d\sigma' \frac{1}{\tau^*(\sigma')} \, P(\sigma',t) 
\end{equation}

\begin{figure}
	[htbp] \centering 
	\includegraphics[width=11cm]{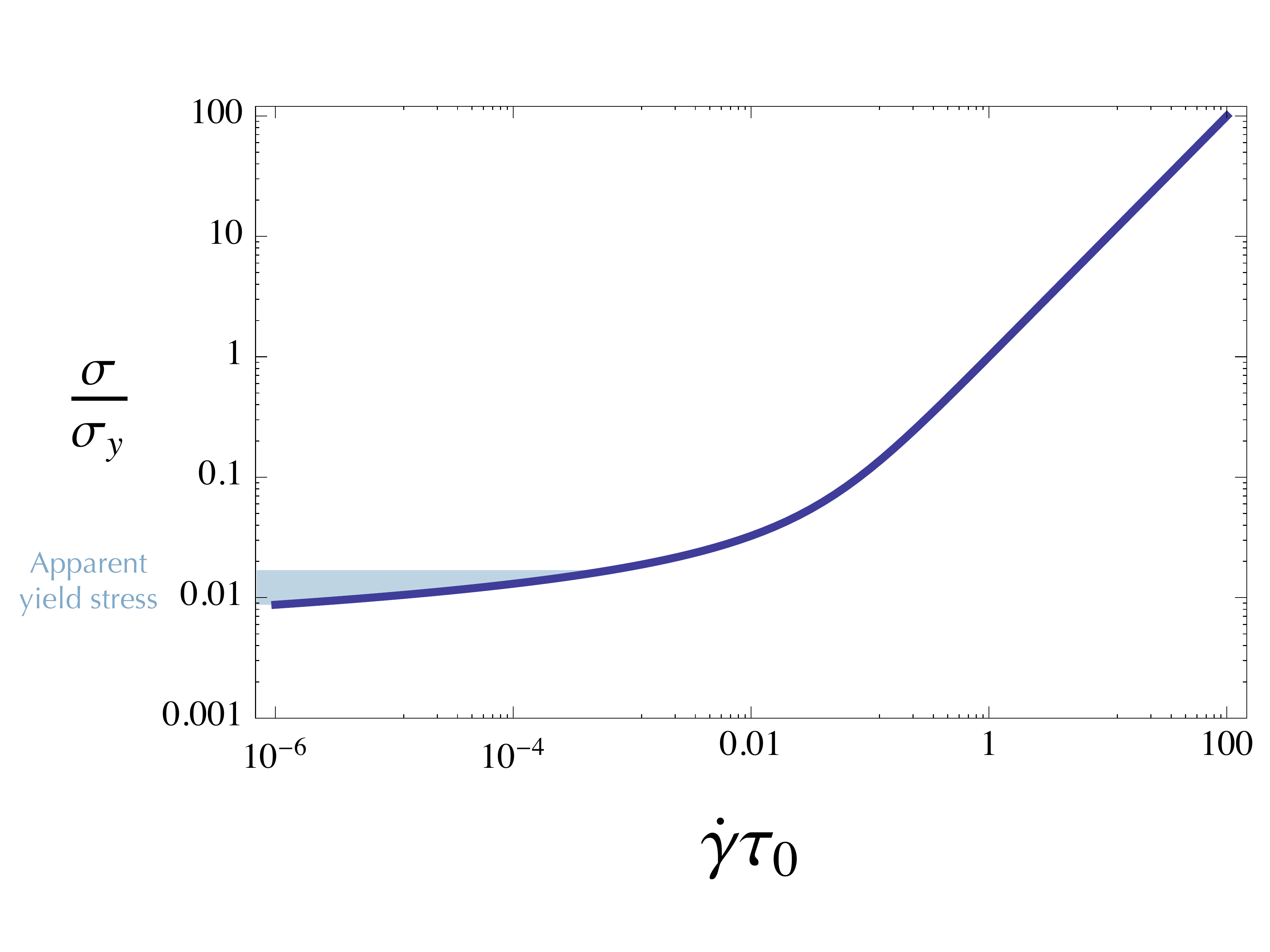} \caption{Flow curve, $\sigma$ as a function of $\dot \gamma$ resulting from equation (\ref{eq:stressEvolution}). The stress $\sigma$ decreases extremely slowly (see text) when $\dot{\gamma} \to 0$: it can be interpreted in terms of an apparent yield stress in experiments and simulations.} \label{fig:rheo} 
\end{figure}

The stationary solution of equation \eqref{eq:stressEvolution} can be computed exactly. It reads $P(\sigma) = C \exp{\left(-\frac{1}{G_P\dot{\gamma}}\int^{\sigma}_{0}\frac{d\sigma'}{\tau^*(\sigma')}\right)}$, with $C$ a normalization constant. From $P(\sigma)$, one can obtain the mean value of the stress in the system and the corresponding flow curve, see Figure~\ref{fig:rheo}. Note that in the limit of small $\dot{\gamma}$, we find back the relation (\ref{flow1}) since the integral over $P(\sigma)$ is dominated by $\sigma'$s such that $\frac{\sigma}{G_P\dot{\gamma}\tau^*(\sigma)}\propto O(1)$. For high shear rates, when the typical distance between nuclei is rather small, a rearrangement of the stress inside one block will change somehow the stresses in the nearby blocks. We have taken this effect into account by modifying the previous equation following the approach developed in \cite{Hebraud:1998,adjari}. The results will be published elsewhere \cite{future-paper}. They lead to a refined description of the flow curve but do not alter our conclusions. Note that for very small shear stresses, the flow is due to the nucleation of just one critical droplet of unstressed material and its subsequent growth until reaching the whole system. In this regime the coarsening-like regime will not be present and the previous description is expected to apply even better. One important conclusion of the analysis of our simple model is that there is no yield stress since the system always flows. However, the ultra-slow logarithmic decay of the stress $\sigma\propto \frac{\sigma_y}{\log (G\dot\gamma)}$ can easily fool the eye and lead to the definition of an effective yield stress when only a limited range of $\dot{\gamma}$ are available, see Fig. \ref{fig:rheo}. Thus, our reasoning has a consequence for the Jamming Diagram of Liu and Nagel~\cite{LiuNagel}, since we have seen that the truly non-flowing phase collapses to three subsets of the $\sigma=0$, the $T=0$, and the $P = \infty$ planes, respectively.

A very intriguing question arises when one considers oscillatory shear with frequency $\omega$. Ferromagnetic systems subjected to an alternating magnetic field have a dynamic phase transition from a low-frequency regime in which nucleation destroys order, to a high frequency regime with spontaneous average magnetisation \cite{Rikvold,Nattermann}. It seems likely that an analogous dynamic phase transition, characterized by the vanishing of the amplitude of the Bragg diffraction peaks, occurs for the flow of solids. One then expects that the density profile of a solid will perform
periodic motion, and would hence stroboscopically stay put,  under periodic shearing throughout a phase of  low amplitude and high frequency.
On the contrary, for frequencies less than $1/\tau^*(\gamma) $, the nucleation mechanism we discussed above will become effective and will lead to a translationally invariant steady state. 
This also applies  to other kinds of solids, such as quasicrystals and ideal glasses (if they exist). Within the flowing phase,
the stroboscopic density, averaged over many cycles, will be spatially flat, while it will have some modulation (quasiperiodic, amorphous) in
the non-flowing regime.  

Our results unveil a general possible mechanism inducing flow of solids (e.g. crystals, quasi-crystals or glasses). Thus they provide upper bounds on the relevant timescales. Whether or not there exist more complicated and faster process is an open question\footnote{For example, for crystals, depending on the way the solid is deformed, the Nabarro-Herring creep mechanism could play an important role.}. In any case, it is interesting to discuss the possibility of observing in experiments the flow phenomena discussed in this work, in particular for three dimensional crystals. Using the value of the kink energy estimated for example for Aluminium, $W\simeq0.075 eV$ \cite{wvalue}, we find $\tau_0\propto 10^{-8}s$ at ambient temperature (working at higher temperatures would diminish this value). Using our previous results, in particular Fig. \ref{fig:rheo}, we find that imposing a stress of the order of $10^{-1}G_P$ lead to deformations of the order of one on timescales of the order of $10^{2}s$. This is clearly a regime that can be investigated in experiments. Usually, real crystals undergo fracture with this kind of values of $\sigma$, but that is because they contain flaws and non-equilibrium defects. The pre-requisite to discover the effect predicted in this work is therefore to prepare perfect crystals. This is not an easy procedure but it can be done \cite{balibar,silicon}. Colloids are other interesting systems to test our results. Micro-rheological experiments in which the motion of particles is visualized by confocal microscopy have been already performed. Actually, it has been possible to study and visualize nucleation of a single dislocation loop \cite{cohen}. This strongly suggests that observing nucleation of the unstressed state and a dynamic phase transition under alternate stress is within reach. The microscopic timescale for colloids is of the order of $ms$; as a consequence, $\tau_0$ becomes much larger and this pushes all timescales roughly 9 orders of magnitude up. Thus to obtain $\dot{\gamma} \propto 10^{-2}$, one should impose a stress of the order of $0.3\,G_P$. This drawback is compensated by two advantages: for hard-sphere colloids, one can rather easily prepare perfect crystals \cite{cohen} and even imposing deformations $\gamma \propto 10^{-1}$ should not lead to fracture. Clearly simulations can also be a very useful tool to unveil the existence of nucleation of unstressed regions inside a stressed solid. Although nucleation is a rare event and, therefore, difficult to obtain in a finite size/time numerical simulation, one could already study the growth of the unstressed regions by starting from an initial condition with the nucleus present. 

Finally, the generalisation of our study to glassy liquids is worth studying, see \cite{mezardyoshino} for a work in this direction. Close to their glass transition they are to a certain extent half-way between solids and liquid. The repeated nucleation of unstressed regions could be relevant to explain rheological properties for small applied shear stress close to the glass transition, where an apparent yield stress emerges.

\begin{acknowledgements}
	We thank S. Balibar, L. Berthier, J.-P. Bouchaud, M. Cates, O. Dauchot, J. Friedel, M. M\'ezard, D.R. Reichman, S. Ramaswamy, G. Tarjus, H. Yoshino for useful discussions. We are grateful to M. M\'ezard and H. Yoshino for discussing with us their unpublished results on micro-rheology of glasses. We thank S. Balibar, L. Berthier, M. M\'ezard, D.R. Reichman and G. Tarjus for feedbacks on the manuscript. GB and FS are partially supported by ANR Grant DYNHET. 
\end{acknowledgements}

\bibliographystyle{spmpsci}

\end{document}